\documentstyle[epsfig]{aipproc}
\def\ra{\rightarrow }

\def\epem{\mbox{e}^+\mbox{e}^- }

\def\kos{\mbox{K}^0_S }
\def\k{\mbox{K}}      
\def\pb{pb$^{-1}$}
\def\NP{{\it Nucl. Phys. }}
\def\PL{{\it Phys. Lett. }}

\def\NIM{{\it Nucl. Inst. Meth. }}
\def\PRep{{\it Phys. Rep. }}
\def\PR{{\it Phys. Rev. }}
\def\PRL{{\it Phys. Rev. Lett. }}
\def\EURO{{\it Eur. Phys. J. }}
\def\ra{\rightarrow } 

\def\epem{\mbox{e}^+\mbox{e}^- }

\def\kos{\mbox{K}^0_S } 
\def\k{\mbox{K}}
\def\pb{pb$^{-1}$} 

\def\q2{\mbox{Q}^2 }

\begin{document}
\title{
       K$^0_s$K$^0_s$ Final State and Glueball Searches\\
       and\\
       $\Lambda\bar{\Lambda}$ Production\\
       in Two-Photon Collisions in L3 at LEP
}

\author{Saverio Braccini$^*$
\footnote{Talk given at Photon2000, Ambleside, England, August 2000.}
}
\address{
$^*$University of Geneva, \\    
24, Quai Ernest Ansermet,  
CH-1211 Geneve 4, Switzerland.\\
E-mail : Saverio.Braccini@cern.ch
}

\maketitle

\begin{abstract}

The $\kos\kos$ final state in two-photon collisions is studied with
the L3 detector at LEP using data collected
at centre-of-mass energies from 91 GeV
to 202 GeV. The mass spectrum is dominated
by the formation of the f$_2\,\!\!\!'$(1525) tensor meson 
in the helicity two state. The two-photon width times the branching
ratio is measured to be 
$\Gamma_{\gamma\gamma}(\mbox{f}_2\,\!\!\!')\times
\mbox{Br(f}_2\,\!\!\!'\ra\k\bar{\k})$= 0.076 $\pm$ 0.006 $\pm$ 0.011 keV.
Clear evidence for destructive f$_2$(1270)-a$_2$(1320) interference is observed.
In addition a clear signal for f$_J$(1750) is observed.
The study of the decay angular distribution in the 1750 MeV mass region
shows that the spin two helicity two wave is dominant. 
No signal is observed in the region around 2.2 GeV. The upper limit for
the two-photon partial width of the $\xi$(2230) tensor glueball candidate of
$\Gamma_{\gamma\gamma}(\xi(2230))\times $Br$(\xi(2230)\ra\kos\kos)<1.4$ eV 
at 95\% C.L. is derived.
The production of $\Lambda \bar{\Lambda}$ pairs in two-photon collisions
is also studied. The cross section is compared to 
quark-diquark model predictions. 

\end{abstract}

\section*{Introduction}

 Electron-positron storage rings are a good laboratory to investigate 
the behaviour of two-photon interactions via the process
$\epem\ra\epem\gamma^*\gamma^*\ra\epem \mbox{X}$, where $\gamma^*$ is a 
virtual photon. 
The outgoing electron and  positron carry nearly  the full 
beam energy and their transverse momenta are usually so small that
they are not detected. In this case the two photons are quasi-real.
This kind of event is 
characterized by an initial state $\epem\gamma^*\gamma^*$, calculable by QED, and
a low multiplicity final state. This process is
particularly useful in the study  of the
formation of hadron resonances and baryon anti-baryon pairs.

The cross section 
is given by the convolution of the QED calculable 
luminosity function $\cal{L}$, giving the flux of
the photons, with the two-photon cross section
$\sigma(\gamma\gamma\ra\mbox{X})$. In the case of the formation of a resonance R,
$\sigma(\gamma\gamma\ra\mbox{R})$ can be expressed by
a Breit-Wigner function. This leads to the
proportionality relation 
$\sigma(\epem\ra\epem\mbox{R})=\cal{K}$ $\Gamma_{\gamma \gamma}(\mbox{R})$
that allows to extract the two-photon width from the cross section.
The quantum numbers of the 
resonance must be compatible with the initial state of the two quasi-real
photons. A neutral, unflavoured meson with even charge conjugation, J$\neq$1
and helicity-zero ($\lambda$=0) or two ($\lambda$=2) can be formed.
In order to decay into $\kos\kos$, a resonance must have
J$^{PC}=(\mbox{even})^{++}$. 
For the $2^{++}$, 1$^3$P$_2$ tensor meson  nonet, the f$_2$(1270), the a$_2^0$(1320) 
and the  f$_2\,\!\!\!'$(1525) can be formed.
However, since these three states are close in mass,
interferences must be taken into account.
According to SU(3), 
the f$_2$(1270) interferes constructively with the a$_2^0$(1320) in the
$\mbox{K}^+\mbox{K}^-$ final state but destructively 
in the $\k^0\bar{\k^0}$ final state~\cite{Lipkin1}.

 Since gluons do not couple directly to photons, 
the two photon width of a glueball is expected to be
very small. A state that can be formed in a gluon rich environment but
not in two photon fusion has the typical signature of a glueball.
According to lattice QCD predictions~\cite{LatticeQCD}, the lowest lying glueball has
J$^{PC}$= 0$^{++}$ and a mass between 1400 and 1800 MeV. 
The 2$^{++}$ tensor glueball is expected in the mass region around 2200 MeV.
Since several 0$^{++}$ states have been observed in the 1400-1800 MeV
mass region, the scalar ground state glueball can  mix with nearby quarkonia,
making the search for the scalar glueball and the interpretation
of the scalar meson nonet a complex
problem~\cite{Close}~\cite{Minkowski}~\cite{Gastaldi}.

 Using the Brodsky-Lepage hard scattering approach\cite{Brodsky-Lepage}, predictions have been 
made for the production of baryon anti-baryon pairs in two-photon interactions. 
The cross section for the process $\gamma \gamma \ra \rm{p} \bar{\rm{p}}$ is measured by
CLEO\cite{CLEO-PP}. Data are found to be
inconsistent with the prediction
of a pure quark model\cite{QQQ} and agreement is found with
the prediction of a quark di-quark 
model\cite{QdQ}, which takes into account non-perturbative quark-quark 
correlations inside the baryon. The cross section $\sigma(\gamma\gamma\ra\Lambda\bar{\Lambda})$
is found by CLEO\cite{CLEO-LL}
to be higher than the value predicted by the quark di-quark model.

 A study of the reactions $\epem\ra\epem\kos\kos$ 
and  $\epem\ra\epem\Lambda \bar{\Lambda}\rm{X}$  is presented here.
The data correspond
to an integrated luminosity of 143 \pb collected by the L3 detector at LEP 
at $\sqrt s=91$ GeV, 52 \pb at $\sqrt s=183$ GeV and
393 \pb at $\sqrt s=189-202$ GeV.
The L3 detector is described in detail elsewhere~\cite{L3}.
These analyses are  mainly based on 
the central tracking system and the high resolution electromagnetic
calorimeter. The events are collected 
predominantly by the charged particle track triggers~\cite{triggers}. 

\section*{$\kos\kos$ final state and glueball searches}

 A study of the reaction $\gamma\gamma\rightarrow$K$^0_s$K$^0_s$
is performed by selecting four charged track events with a net charge
of zero and 
$|\vec{p_T}(\pi^+\pi^-\pi^+\pi^-)|^2<0.1$ GeV$^2$.
Within these events two $\kos$  candidates are searched for 
by secondary vertex reconstruction. 
Events with photons are rejected.
\begin{figure}[t] 
\centerline{
\epsfig{file=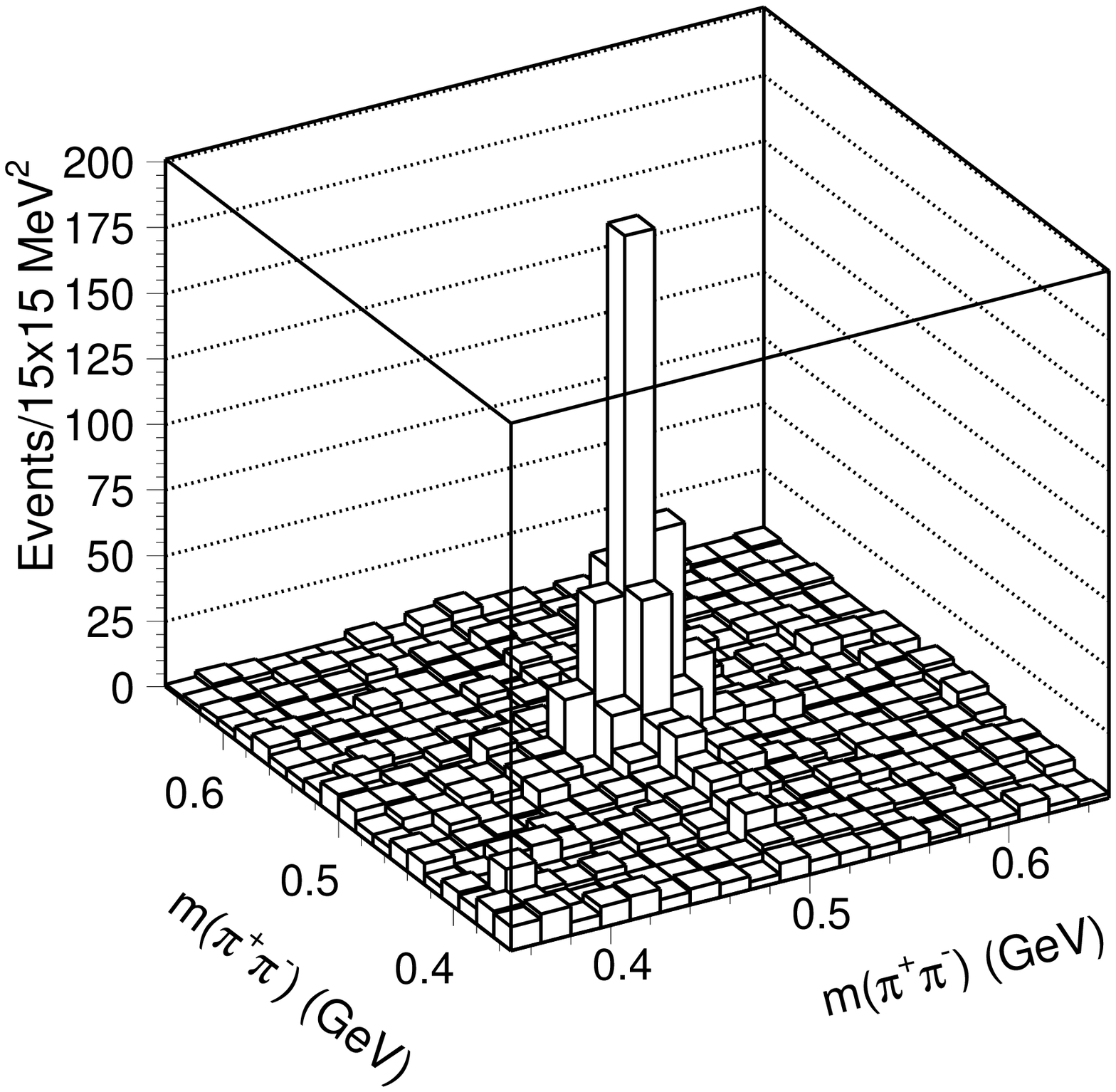,height=3.5in,width=3.5in}
\hspace*{-1.0cm}
\epsfig{file=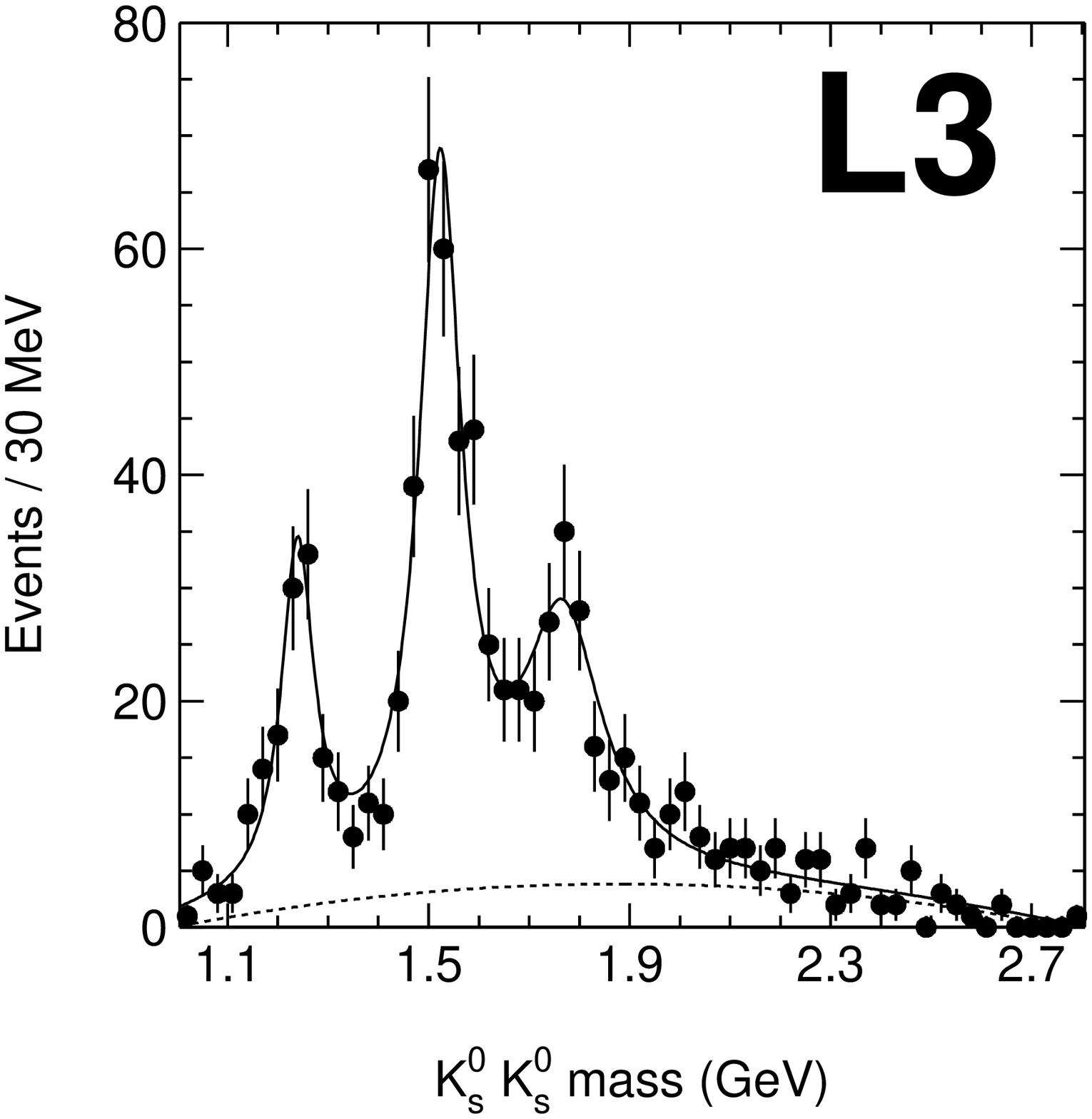,height=2.8in,width=2.8in}
}
\hspace{10pt}
\caption{
(a) The mass of a $\kos$ candidate versus the mass of the other candidate.
(b) The $\kos\kos$ mass spectrum.
}
\label{fig1}
\end{figure}
 Fig.~\ref{fig1}(a) shows the distribution of the mass of
one $\kos$ candidate versus the mass of the other candidate.
There is a strong enhancement corresponding to the
$\kos\kos$ exclusive formation over a small background. We require
that the invariant masses of the two $\kos$ candidates must be inside
a circle of 40 MeV radius centred on the peak of the 
$\kos\kos$ signal.
With these selection criteria, 802 events are found in the data sample.
 The resulting $\kos\kos$ mass spectrum is shown in 
fig.~\ref{fig1}(b). The spectrum is dominated by the formation of the
f$_2\,\!\!\!'$(1525) tensor meson. The mass region between 1100 and 1400 MeV
shows destructive f$_2(1270)-$a$_2^0(1320)$
interference in  the $\kos\kos$ final state~\cite{Lipkin1}. 
A clear signal is visible for the f$_J$(1750).
No excess is observed around 2230 MeV.
 A maximum likelihood fit using three Breit-Wigner 
functions plus a second order polynomial for the background is 
performed on the full $\kos\kos$ mass spectrum. 
The fit is shown in fig.~\ref{fig1}(b) and the results are summarized in 
Table~\ref{table:fit}. 

\begin{table}[b]
\caption{Results from the maximum likelihood fit on the full $\kos\kos$ mass spectrum.}
\label{table:fit}
\begin{tabular}{cccc}
\rule[.15in]{0.0in}{0.0in}&f$_2$(1270)-a$_2$(1320)  & f$_2\,\!\!\!'$(1525)  & f$_J$(1750)\\ \hline
Mass  (MeV)         & 1239  $\pm$   6  & 1523 $\pm$  6    & 1767 $\pm$ 14   \\ 
Width (MeV)         &   78  $\pm$  19  &  100 $\pm$ 15    &  187 $\pm$ 60   \\ 
Area                &  123  $\pm$  22  &  331 $\pm$ 37    &  220 $\pm$ 55   \\ 
\end{tabular}
\end{table}

 In order to correct the data for the detector acceptance and efficiency, 
a Monte Carlo procedure is used~\cite{Linde}. 
The nominal f$_2\,\!\!\!'$(1525) parameters~\cite{PDG} 
are used for the generation. 
The angular distribution of the two $\kos$'s in the two-photon center-of-mass system
is generated according to
phase space (J=0) i.e. uniform in $\cos\theta^*$ and in $\phi^*$, where
$\theta^*$ and $\phi^*$ are the polar and azimuthal angles taking
the $z$ direction parallel to the electron beam.
In order to take into account the helicity of a spin-two resonance, 
a weight is assigned to each generated event according to the 
weight functions:  
$w=(\cos^2\theta^*-\frac{1}{3})^2$ for the spin-two helicity-zero (J=2, $\lambda$=0) contribution and
$w=\sin^4\theta^*$ for the spin-two helicity-two (J=2, $\lambda$=2) contribution.
All the events are passed through the full detector simulation program 
and are reconstructed following the same procedure used for the data.

\begin{figure}[t] 
\centerline{
\hspace*{.3cm}
\epsfig{file=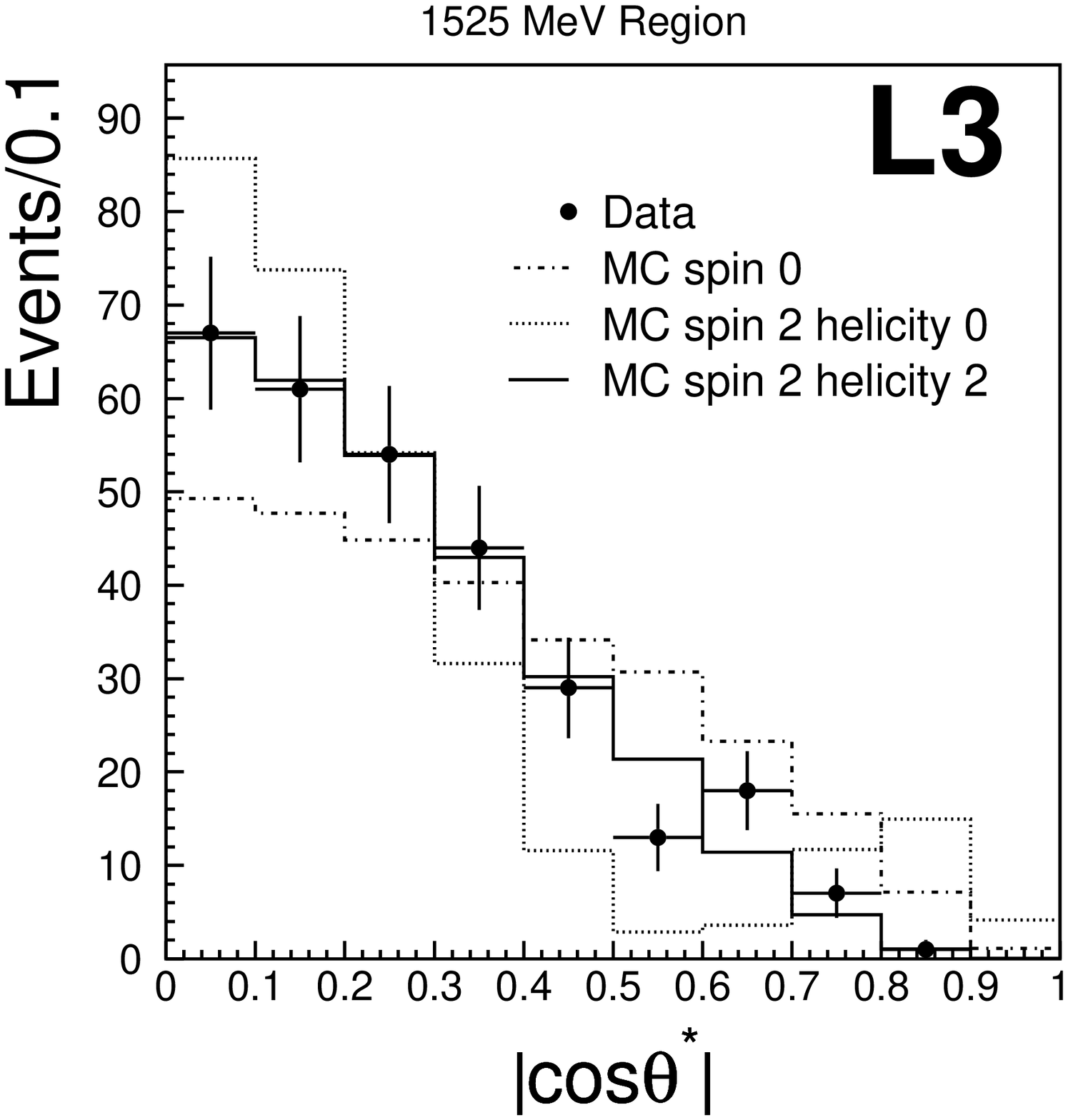,height=3.0in,width=3.0in}
\epsfig{file=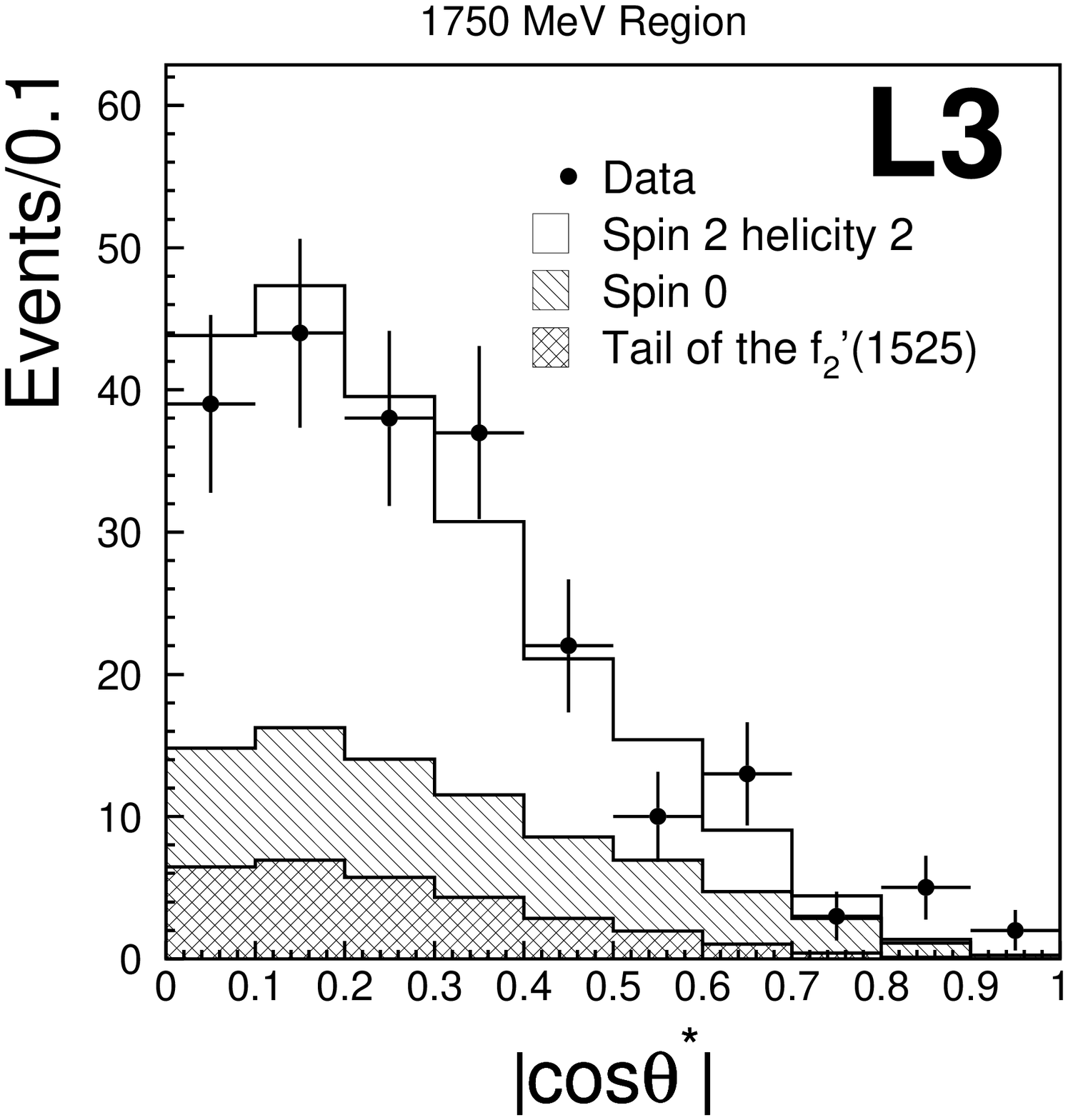,height=3.0in,width=3.0in}
}
\vspace{10pt}
\caption{
(a) The experimental angular distribution is compared with 
different spin-helicity
Monte Carlo
predictions for the f$_2\,\!\!\!'$(1525).
(b) The result of the fit of the angular distribution in the 1750 MeV mass region.
}
\label{fig2}
\end{figure}
 
 To determine the spin and the helicity state in the
mass region of the f$_2\,\!\!\!'$(1525), 
a study of the angular distribution of the two $\kos$'s
in the mass region between 1400 and 1640 MeV in the
two-photon center of mass is performed.
The experimental polar angle distribution is compared with the
Monte Carlo  in 
Fig.~\ref{fig2}(a) for pure (J=0),
(J=2, $\lambda$=0) and (J=2, $\lambda$=2) cases.
A $\chi^2$ is calculated normalizing the
Monte Carlo distributions to the same number of events as in the data.
Bins are grouped in order to have at least 10 entries both in the data
and in the Monte Carlo.
The confidence level values for the (J=0) and (J=2, $\lambda$=0) 
hypotheses are less then $10^{-6}$. For the
(J=2, $\lambda$=2) hypothesis a confidence level of 99.9\%
is obtained. The contributions of (J=0) and (J=2, $\lambda$=0) 
are found to be compatible with zero when fitting the three waves
simultaneously. 
Thus data are in agreement with a pure spin
two, helicity two wave contribution.
According to  the theoretical predictions~\cite{Kopp}, the
(J=2, $\lambda$=2) contribution largely dominates over (J=2, $\lambda$=0).
From the cross section the value 
$\Gamma_{\gamma\gamma}(\mbox{f}_2\,\!\!\!'(1525))\times
\mbox{Br(f}_2\,\!\!\!'(1525)\ra\k\bar{\k})$= 0.076 $\pm$ 0.006 $\pm$ 0.011 keV is
obtained for the two-photon width. The main source of
systematic error is due to the background parameterization in the fitting procedure.
This result is in good agreement with the value previously published by L3~\cite{Saverio}.
 
  To investigate the spin composition in the mass region between 1640 and 2000 MeV 
the angular distribution of the two $\kos$'s
in the two-photon center of mass is studied. 
A resonance with a mass of 1750 MeV and
a total width of 200 MeV is generated with the same Monte Carlo procedure
adopted for the f$_2\,\!\!\!'$(1525). 
The effect of the tail of the f$_2\,\!\!\!'$(1525) in this mass region is
taken into account by using the Monte Carlo distribution for the f$_2\,\!\!\!'$(1525) with
(J=2, $\lambda$=2). The fraction of the events belonging to the f$_2\,\!\!\!'$(1525)
in the 1750 MeV mass region is found to be 14\%.
 A fit of the angular distribution is performed using a combination of the two
waves (J=0) and (J=2, $\lambda$=2) for the signal plus the
distribution of the tail of the f$_2\,\!\!\!'$(1525). The contribution due to
the (J=2, $\lambda$=0) wave is considered negligible with respect to (J=2, $\lambda$=2)
according to 
the theoretical predictions~\cite{Kopp} and to our experimental results for the 
f$_2\,\!\!\!'$(1525).
The normalization is fixed to
the same number of events as in the data and
bins are grouped in order to have at least 10 entries both in the data
and in the Monte Carlo. The fit is shown in Fig.~\ref{fig2}(b). 
The fraction of (J=0) is found to be
24$\pm$16\%. The confidence level is  68\%. Thus the (J=2, $\lambda$=2) wave 
is found to be dominant also in
the 1750 MeV mass region. If the (J=2, $\lambda$=0) wave is also considered,
its contribution is  found to be compatible with zero.

 The BES Collaboration~\cite{BESKPKM} reported the presence of both 2$^{++}$ and 0$^{++}$
waves in the 1750 MeV mass region in K$^+$K$^-$ in the
reaction $\epem\ra\mbox{J}/\psi\ra\mbox{K}^+\mbox{K}^-\gamma$. The (J=0) fraction is
estimated to be 30$\pm$10\%, in good agreement with our measurement.

 For the two-photon width of the (J=2, $\lambda$=2) state, we measure
$\Gamma_{\gamma\gamma}(\mbox{f}_2(1750))\times
\mbox{Br(f}_2(1750)\ra\k\bar{\k})$= 0.049 $\pm$ 0.011 $\pm$ 0.013 keV.
The systematic error takes into account the selection criteria, the trigger,
the fitting procedure, the uncertainty on the total width and on the
(J=2, $\lambda$=2) fraction.

 The (J=2, $\lambda$=2) signal around 1750 MeV may be due to the formation of a radially
excited tensor meson state.
The f$_2''$ is in fact predicted to have a mass of 1740 MeV
and a two-photon width of 1.040 keV~\cite{Munz}.
 The 1750 MeV region is also very interesting for glueball searches and for
the understanding of the 0$^{++}$ scalar meson nonet. 
In the scenario presented in~\cite{Close}, the f$_0$(1500) is identified with
the scalar  0$^{++}$ ground state glueball.
This hypothesis includes the prediction of
a further scalar state, the f$_0\,\!\!\!'$(1500-1800), mainly composed by $s\bar{s}$. This state
will couple strongly to K$\bar{\mbox{K}}$ and its observation is essential to support
the f$_0$(1500) glueball nature. Our and BES measurements support this hypothesis.

 The BES Collaboration confirmed the observation by the Mark III Collaboration
of a resonance, 
the $\xi$(2230)~\cite{xi2230}, produced in radiative decays of the $\rm{J}/\psi$. 
Due to its narrow width and its production in gluon rich environment,
this state is considered a glueball candidate.
Its mass is consistent with the lattice QCD prediction for the
ground state tensor glueball. 

 Since we do not observe any signal in the 2230 MeV mass region, 
a Monte Carlo simulation is used to
determine the detection efficiency for the $\xi$(2230)
under the hypothesis of a pure (J=2, $\lambda$=2) state.
For the simulation we use
a mass of  2230 MeV and a total width of 20 MeV. A mass resolution
of $\sigma$= 60 MeV is found. 
The signal region is chosen to be $\pm2\sigma$ around the 
$\xi$(2230) mass. In order to evaluate the background two sidebands of $2\sigma$ are considered.
Using the standard method~\cite{PDG} for extracting 
an upper limit for a Poisson distribution with background, we determine the upper 
limit 
$\Gamma_{\gamma\gamma}(\xi(2230))\times $Br$(\xi(2230)\ra\kos\kos)<1.4$ eV 
at 95\% C.L.

 Since gluons do not couple directly  to photons, the two-photon width
is expected to be small for a glueball. To make this statement more quantitative, a parameter called
stickiness~\cite{Chanowitz} is introduced. It is expected to be of the order of one for
quarkonia and much larger for glueballs.
 Combining the results reported by BES and Mark III for the 
$\Gamma(\mbox{J}/\psi\rightarrow\gamma\xi(2230))\times $Br$(\xi(2230)\ra\kos\kos)=$ 1.9$\pm$0.5 eV
and our upper limit on $\Gamma(\xi(2230)\rightarrow\gamma\gamma)\times $Br$(\xi(2230)\ra\kos\kos)$, 
we obtain
a lower limit on the stickiness $S_{\xi(2230)} >$73 at 95\% C.L. 
This value is in agreement with the measurements by CLEO~\cite{CLEO-KK} and
is much larger than
the values measured for all the well established $\rm{q}\bar{\rm{q}}$ states and supports the interpretation
of the $\xi$(2230) as the tensor glueball. 
A further confirmation of its existence 
in gluon rich environments becomes now very important.

\section*{$\Lambda\bar{\Lambda}$ production}

\begin{figure}[t] 
\centerline{
\epsfig{file=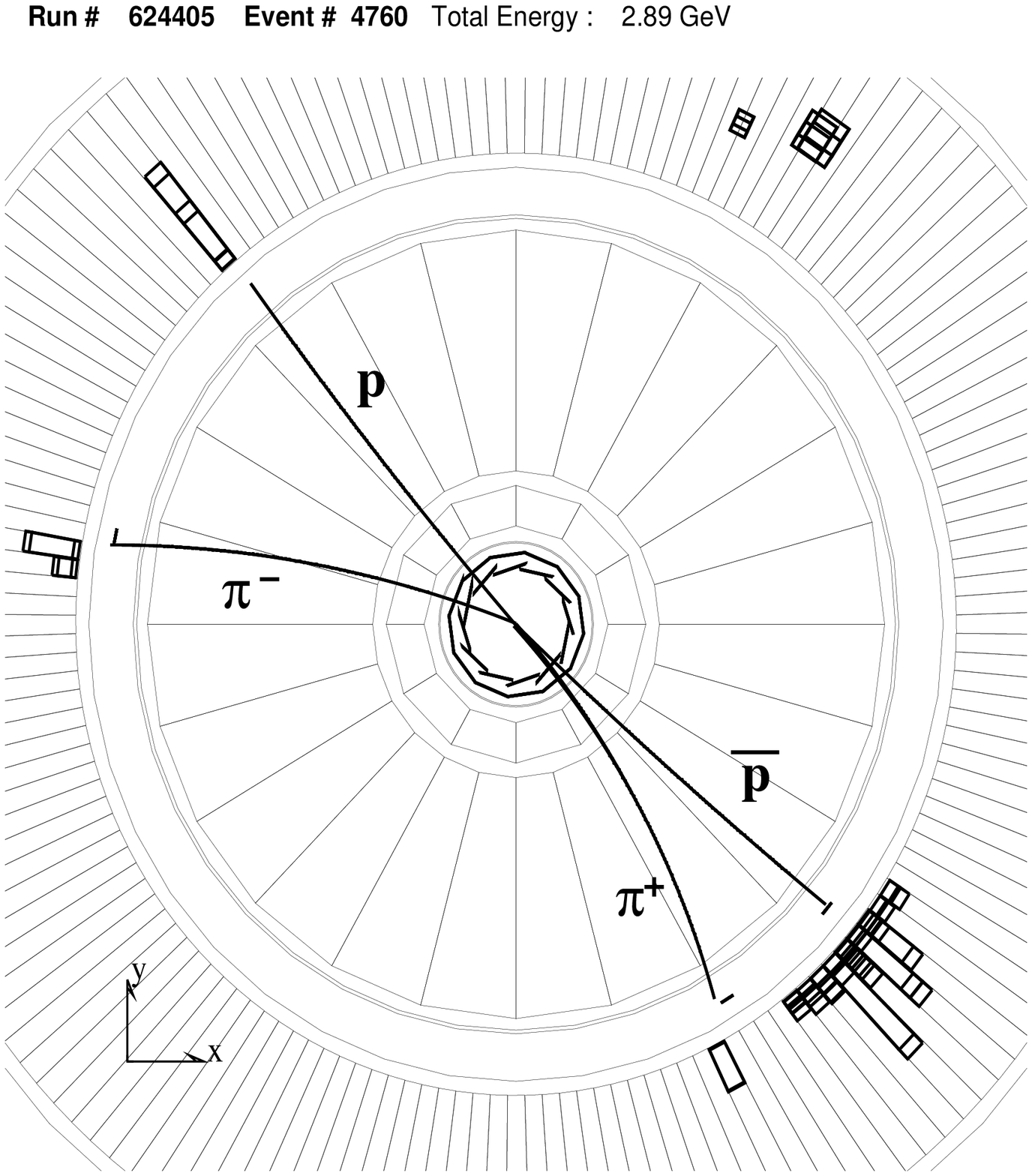,height=2.7in,width=2.7in}
\epsfig{file=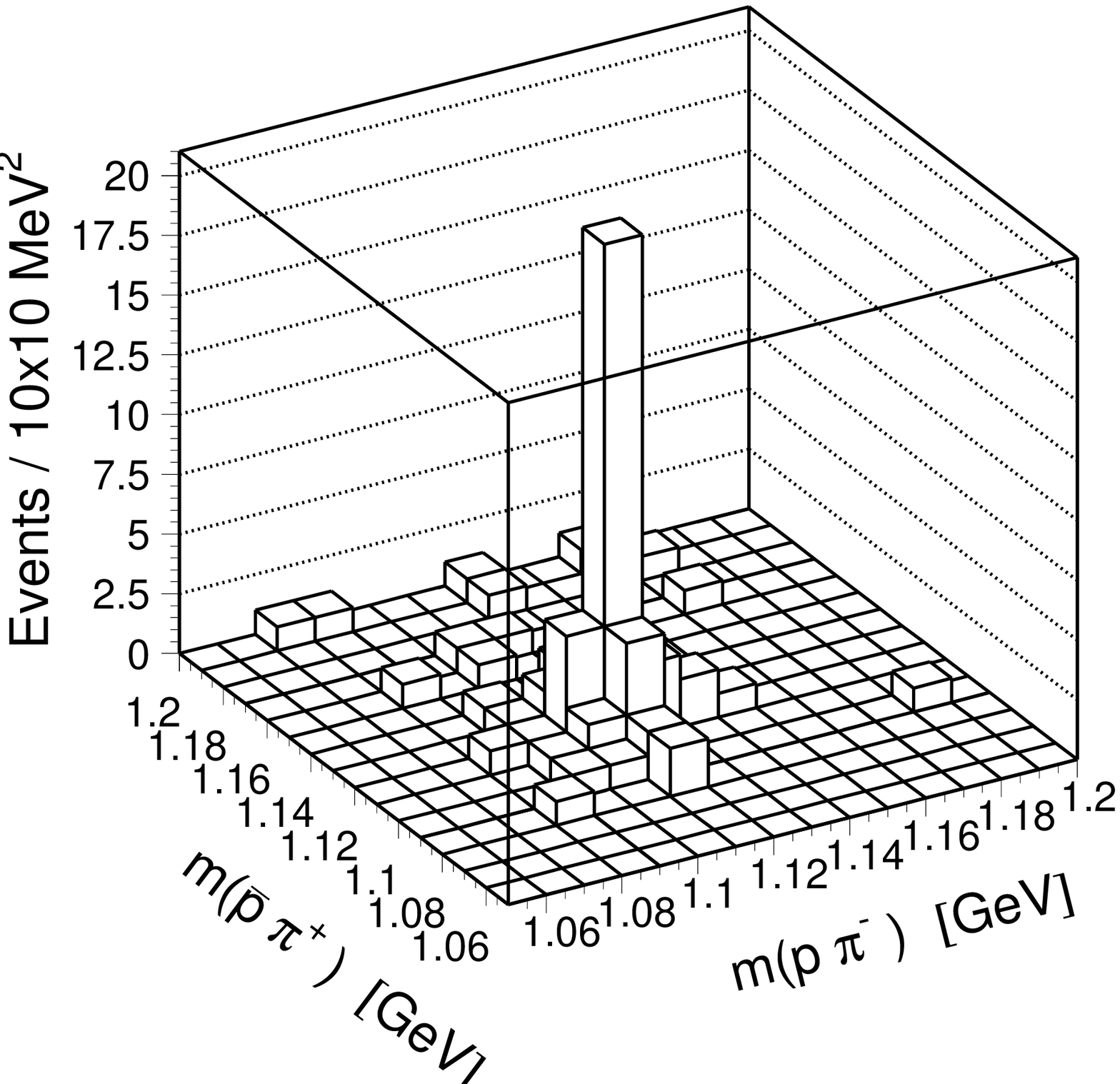,height=3.1in,width=3.1in}
}
\vspace{10pt}
\caption{
(a) A typical event $\epem\ra\epem\Lambda\bar{\Lambda}\mbox{X}$.
(b) The signal of $\Lambda\bar{\Lambda}$ production.
}
\label{fig3}
\end{figure}

 The production of $\Lambda\bar{\Lambda}$ pairs in two-photon collisions
is studied using the
decays $\Lambda\rightarrow \rm{p} \pi^-$ and $\bar{\Lambda}\rightarrow \bar{\rm{p}} \pi^+$.
Events with four charged tracks, a net charge of zero and two secondary 
vertices are selected. For each secondary vertex the mass of the proton is
assigned to the track with the largest transverse momentum. The ionization loss measurements
in the tracking chamber must be consistent with the 
$\rm{p} \pi^- \bar{\rm{p}} \pi^+$ hypothesis. The track of the
anti-proton candidate must be in correspondence with a large deposit
of energy ($E_{em}>0.35$ GeV) in the electromagnetic calorimeter
produced by its annihilation. A typical event is shown in fig.~\ref{fig3}(a) 
where the $\Lambda\bar{\Lambda}$ topology and in particular 
the annihilation of the anti-proton is clearly visible.
In fig.~\ref{fig3}(b) the mass distribution of the $\Lambda$ candidate
is shown versus the mass of the $\bar{\Lambda}$ candidate: the signal due to 
$\Lambda\bar{\Lambda}$ production is very clear and 44 events are selected inside
a circle of 40 MeV radius around the peak of the
$\Lambda\bar{\Lambda}$ signal.

\begin{figure}[t] 
\centerline{
\hspace*{.8cm}
\epsfig{file=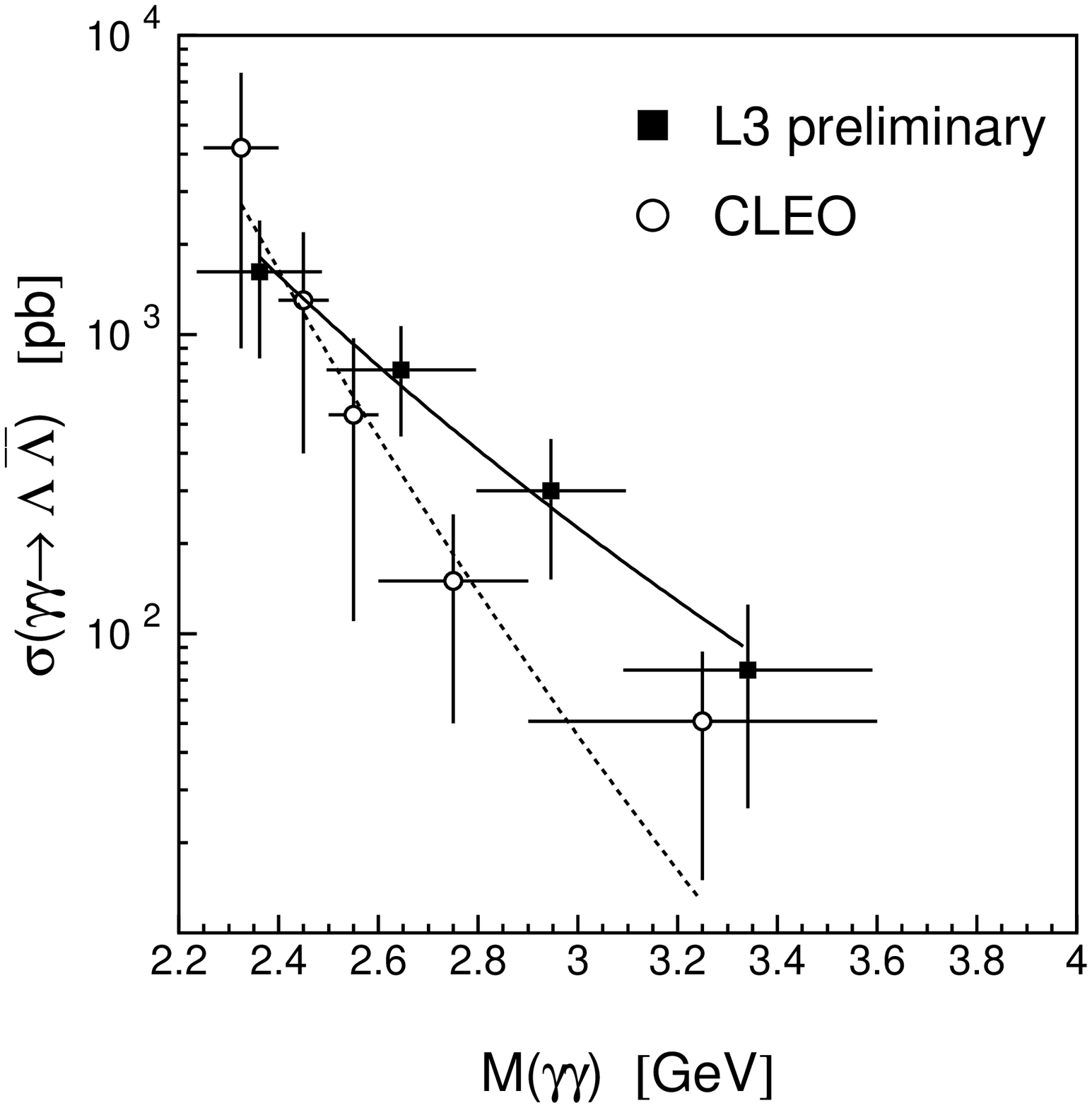,height=3.1in,width=3.1in}
\hspace*{-.5cm}
\epsfig{file=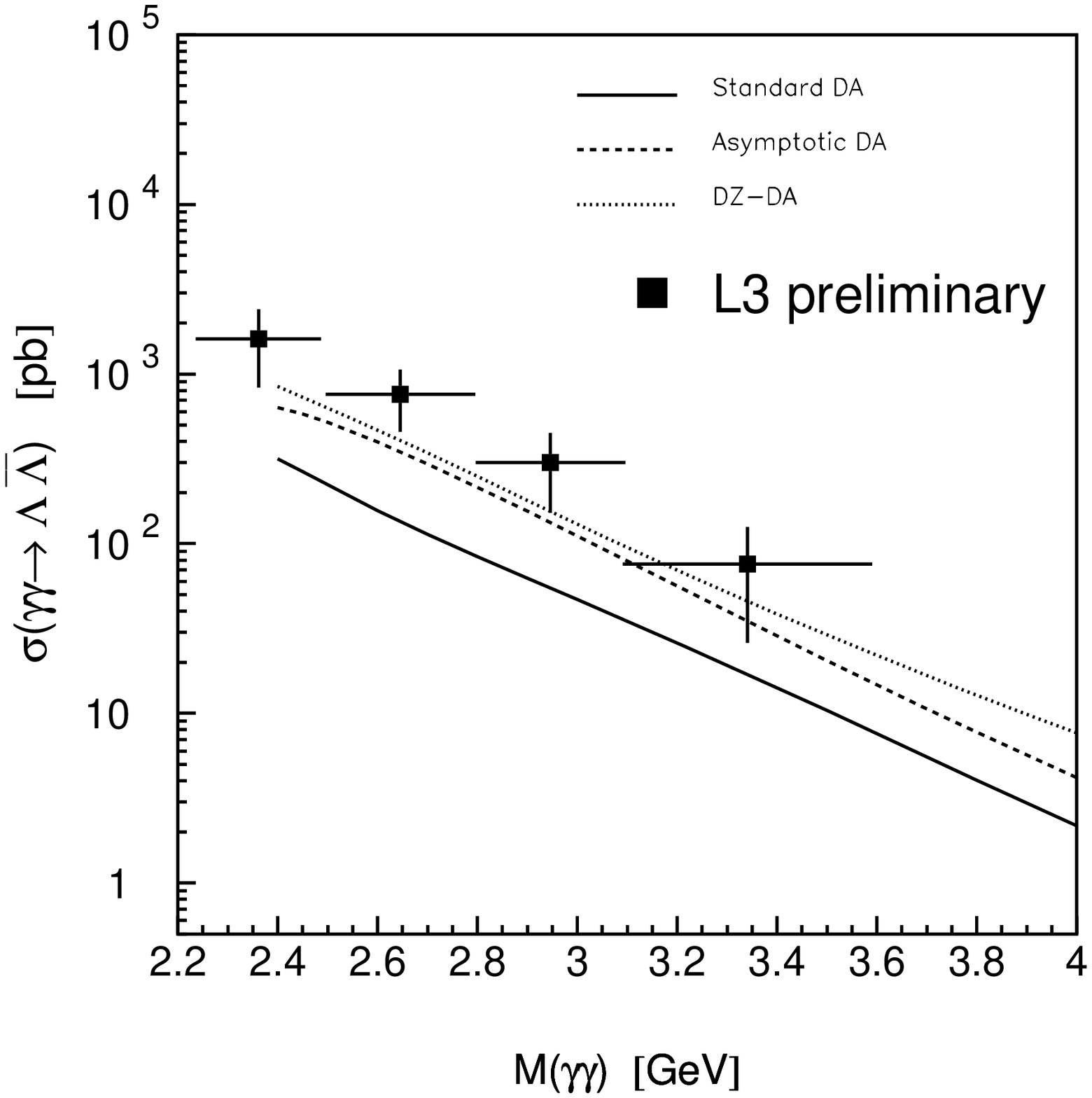,height=3.1in,width=3.1in}
}
\vspace{10pt}
\caption{
The measured cross section
$\sigma(\gamma\gamma\rightarrow\Lambda\bar{\Lambda}\rm{X})$
is compared to CLEO data (a) and to quark di-quark model theoretical predictions (b).
}
\label{fig4}
\end{figure}

 Since no cut is applied to
$|\vec{p_T}(\rm{p}\pi^-\bar{\rm{p}}\pi^+)|^2$ 
and events with
photons are not removed, the inclusive cross section 
$\sigma(\gamma\gamma\rightarrow\Lambda\bar{\Lambda}\rm{X})$
is measured by deconvoluting the luminosity function. The result
is shown in fig.~\ref{fig4}(a) together with
CLEO~\cite{CLEO-LL} data. The two measurements are in agreement within the large errors. 
Fig.\ref{fig4}(a) shows also a fit of the form $\sigma\propto M^{-n}$ that gives
$n=8.7\pm 3.8$ for our data and $n=16.1\pm 5.8$ for CLEO. According to the
dimensional counting rule~\cite{Brodsky}, the value of $n$ is 12 for a baryon composed
by three quarks and 8 for a baryon composed by a quark and a di-quark. Thus
our data favor the quark di-quark model.

 In fig.\ref{fig4}(b) our measurement is compared to recent calculations 
of
$\sigma(\gamma\gamma\rightarrow\Lambda\bar{\Lambda})$
in the
framework of the quark di-quark model~\cite{Berger} for three different distribution
amplitudes for the octet baryons. The mass dependence is in good agreement with our data
that lie above the predictions. The excess in the data may be due to the fact that
the contribution of $\Sigma^0$, $\bar{\Sigma^0}$ and other baryons is not removed. 

\section*{Acknowledgments}

 I would like to acknowledge all the members of the two-photon physics
analysis group of the L3 Collaboration, in particular B. Echenard,
J.H. Field, M.N. Focacci-Kienzle and M. Wadhwa.


\end{document}